\documentclass[12pt]{article}
\usepackage{epsfig}

\title{Parallel tempering and decorrelation of topological charge in full QCD}

\author{E.-M.~Ilgenfritz$\,^a$, W.~Kerler$\,^b$,
        M.~M\"uller-Preussker$\,^b$, H.~St\"uben$\,^c$}
\date{{}}

\newcommand{\WIDTH}{4.5cm}  
\newcommand{\Width}{0.3cm}  
\newcommand{\qTEXT}[1]{\makebox[\WIDTH][c]{\small\textbf{#1}}}
\newcommand{\qtext}[1]{\makebox[\WIDTH][c]{\small #1}}
\newcommand{\qpic}[1]{\epsfig{file=q.#1.eps,width=\WIDTH,%
bbllx=0,bblly=0,bburx=360,bbury=220}}

\begin{document}

\maketitle

\vspace*{-6.7cm}\begin{flushright}{\small 
HU--EP--00/31\\
RCNP--Th00027\\
ZIB--Report 00--25 
}\end{flushright}\vspace*{3.6cm}

\begin{center}
{\sl\noindent $^a$ Research Center for Nuclear Physics Osaka University, 
Japan\\
$^b$ Institut f\"ur Physik, Humboldt Universit\"at, D-10115 Berlin, Germany\\
$^c$ Konrad-Zuse-Zentrum f\"ur Informationstechnik, D-14195 Berlin, Germany}
\end{center}

\vspace*{4mm}

\begin{abstract}
The improvement of simulations of QCD with dynamical Wilson fermions
by combining the Hybrid Monte Carlo algorithm with parallel tempering
is studied on $10^4$ and $12^4$ lattices.  As an indicator for
decorrelation the topological charge is used.
\end{abstract}

\vspace*{1mm}

\section{Introduction}

Decorrelation of the topological charge in Hybrid Monte Carlo (HMC)
simulations of QCD with dynamical fermions is a long standing problem.
For staggered fermions an insufficient tunneling rate of the topological 
charge $Q$ has been observed \cite{MMP,Pisa}. For Wilson fermions the 
tunneling rate has been reported to be adequate in many cases
\cite{SESAM,CP-PACS}. However, since the comparison is somewhat subtle,
the reason for this could also be that one is not as far in the critical 
region as with staggered fermions. In any case for Wilson fermions on large 
lattices and for large values of $\kappa$ near the chiral limit the 
distribution of $Q$ is not symmetric even after more than 3000 trajectories 
(see e.g.\ Figure~1 of \cite{SESAM}).

While many observables appear not sensitive to topology there are clearly
ones for which proper account of topological sectors is important. In fact,
it has been observed that the $\eta'$ correlator is definitely $Q$ dependent 
\cite{SESAM2}. Thus it is quite important to look for simulation methods that 
produce good distributions of $Q$. 

Generally one wants to penetrate as deeply as possible into the critical
region. This, however, is limited by the increasing autocorrelation times.
Thus better decorrelation is also desirable with respect to this goal. 
The topological charge appears to be a good touchstone
of the improvements achieved by new methods.

The method of \emph{simulated} tempering has been proposed in
\cite{mar92} where the inverse temperature has been made a dynamical
variable in the simulations.  The principle is, however, much more
general in the sense that any parameter in the action can be made
dynamical. The mechanism behind the better decorrelation of the simulation 
process is that in place of suppressed tunneling an easy detour through 
parameter space with little suppression is opened.

In fact, considerable improvements have been obtained by simulated tempering
with a dynamical number of the degrees of freedom in the Potts-Model 
\cite{ker93}, with a dynamical inverse temperature for spin glass \cite{ker94} 
and with a dynamical monopole coupling in U(1) lattice theory \cite{ker95}. 
An investigation with a dynamical mass of staggered fermions in full QCD 
\cite{boy97} has indicated a potential gain at smaller masses. Simulated 
tempering requires the determination of a weight function in the generalized 
action. Thus developing efficient methods \cite{ker94,ker95} for obtaining the 
weight function has been a crucial issue. 

A major progess was the proposal of the \emph{parallel} tempering method 
\cite{HN}, in which no weight function needs to be determined. This has allowed
large improvements in the case of spin glass \cite{HN}. In QCD also
improvements have been obtained with staggered fermions \cite{Boyd}.
On the other hand, in simulations of QCD with O($a$)-improved Wilson
fermions \cite{UKQCD} no gain has been found. However, the use of only two 
ensembles in this study did not take advantage of the main idea of the method. 

In the present paper parallel tempering is used in conjunction with
HMC to simulate QCD with (standard) Wilson fermions. To study the
performance of the tempering method we have recorded the time series
of the topological charge. In previous work we have done this on an
$8^4$ lattice \cite{us} where already a considerable increase of the
transitions of the topological charge was observed.  However, the effect
of enlarging the lattice size remained a major question. Furthermore, the 
rather narrow distribution of the topological charge for this lattice size 
prevented us from resolving more details. Therefore, investigations 
on $10^4$ and $12^4$ lattices are done here.

\section{Parallel tempering}

In standard Monte Carlo simulations one deals with one parameter set 
$\lambda$ and generates a sequence of configurations $C$. The set $\lambda$ 
includes $\beta$, $\kappa$ and as technical parameters the leapfrog time step 
and the number of time steps per trajectory. $C$ comprises the gauge field and 
the pseudo fermion field.

In the parallel tempering approach \cite{HN,EM} one simulates $N$ 
ensembles $(\lambda_i; C_i),\,i = 1, \dots, N$ in a
single combined run.  
The whole run represents one bigger ensemble in an enlarged configurations 
space stratified (with respect to $\lambda$) into the ensembles above.
Two steps alternate: (a) update of $N$
configurations in the standard way, (b) exchange of configurations by
swapping pairs. Swapping of a pair of configurations means
\begin{equation}
((\lambda_i; C_i), (\lambda_j; C_j)) \rightarrow \left\{
\begin{array}{ll}
\!((\lambda_i; C_j), (\lambda_j; C_i)), & \!\mbox{if accepted} \\
\!((\lambda_i; C_i), (\lambda_j; C_j)), & \!\mbox{else}
\end{array}
\right.
\end{equation}
with the Metropolis acceptance condition 
\begin{equation}
P_{\rm swap}(i,j) = \min\left( 1, e^{-\Delta H} \right) \,,
\label{Pswap}
\end{equation}
\begin{equation}
\Delta H = 
H_{\lambda_i}(C_i) + H_{\lambda_j}(C_j) -
H_{\lambda_i}(C_j) - H_{\lambda_j}(C_i)
\end{equation}
where $H$ is the Hamiltonian of the HMC dynamics.
Since after swapping due to detailed balance both ensembles remain in any case 
in equilibrium \cite{HN,EM}, the swapping sequence can be freely chosen. In 
order to achieve a high swap acceptance rate one will only try to swap 
$(\beta,\kappa)$-pairs that are close together. If the chosen 
$(\beta,\kappa)$-values lie on a curve in the $(\beta,\kappa)$-plane there are 
three obvious choices for the swapping sequence of neighboring 
$(\beta,\kappa)$-pairs. One can step through the curve in either direction or 
swap randomly. We have observed that it is advantageous to step along such a 
curve in the direction from high to low tunneling rates of $Q$.

\section{Simulation details}

Wilson fermions and the standard one-plaquette action for the gauge fields
have been used. The HMC program applied the standard conjugate gradient
inverter with even/odd preconditioning. The trajectory length was
always~1. The time steps were adjusted to get acceptance rates of
about 70\%. In all cases 1000 trajectories were generated (plus
50--100 trajectories for thermalization).

$Q$ was measured using its naively discretized plaquette form after doing
50 cooling steps of Cabibbo-Marinari type. This method gives close to integer 
values which were rounded to the nearest integers.

Statistical errors were obtained by binning, i.e., the values given are
the maximal errors calculated after blocking the data into bins of
sizes $10,20,50$ and $100$.

\section{Results}

We have performed tempered runs using 6 and 7 ensembles, all at $\beta = 5.6$ 
on $10^4$ and $12^4$ lattices, as well as standard HMC runs at fixed $\kappa$ 
values for comparison. Our ensembles cover the $\kappa$-range investigated by 
SESAM ($\kappa = 0.156, 01565, 0.157, 0.1575$) \cite{SESAM}. 
In the run using 6 ensembles we studied the effect of extending the 
$\kappa$-range by adding lower values of $\kappa$, while in the run using 7 
ensembles we have tested the efficiency of a denser spacing of the 
$\kappa$-values. Our $\kappa$-values are listed in Table \ref{t1}. 

Figures 1 and 2 show time series of $Q$ obtained on $10^4$ and $12^4$ 
lattices, respectively, for standard HMC and for tempered HMC with 6 and 
with 7 ensembles.  One sees that tempering makes $Q$ fluctuating much 
stronger. Thus correlations between subsequent trajectories indeed decrease.
Comparing the time series of $Q$ on $10^4$ and $12^4$ lattices, the increasing 
width of the topological-charge distribution is seen to lead to 
a richer pattern of transitions.

On the $10^4$ lattice, in addition to the tempered run with 7 ensembles with 
finer steps in $\kappa$ and the one with 6 ensembles which penetrates deeper 
into the uncritical region we also performed a tempered simulation which
used only the four common $\kappa$-points of the latter ones. The results
of this indicate that both of these modifications cause only moderate
improvements, which suggests that optimizing range and distances one might do 
with fewer ensembles. 

In principle an autocorrelation analysis would allow to quantitatively assess 
the improvement of decorrelating. However, such an analysis cannot 
be carried out with the given size of samples. Some quantitative account is 
neverless possible considering the mean absolute change of $Q$, called 
mobility in \cite{SESAM},
\begin{equation}
M_Q = \frac{1}{N_{\rm traj}} \sum_{i=1}^{N_{\rm traj}}
\left| Q(i) - Q(i - 1) \right|
\label{D1}
\end{equation}
and the HMC time between topological events $T_Q$ given by  
\begin{equation}
1/T_Q = \frac{1}{N_{\rm traj}} \sum_{i=1}^{N_{\rm traj}}
\min \left(1, 
\left| Q(i) - Q(i - 1) \right| \right)\;.
\label{M1}
\end{equation}
While $M_Q$ measures the change in topological charge, $T_Q$ tells how
often changes (without regard to their magnitude) occur. The difference between 
$M_Q$ and $1/T_Q$ increases with the width of the distributions of the 
topological charge.

The results obtained for $M_Q$ and $1/T_Q$ are given in Table \ref{t1}. 
Comparing $M_Q$ for standard and tempered runs, gains by factors 2 to 5 are 
obvious. At larger $\kappa$ values the improvement in the tempered run with 7 
ensembles is larger than in that with 6 ensembles. Thus using finer steps in 
$\kappa$ pays off more than including additional $\kappa$-values in the
uncritical region. Comparing $M_Q$ and $1/T_Q$ a significant effect of the 
broader $Q$-distribution can be seen. As expected, this effect becomes stronger
on the larger lattice. Generally one observes that both $M_Q$ and $1/T_Q$ 
increase with the lattice size.

The results obtained for $\langle Q\rangle$ and for $\langle Q^2\rangle$ are 
given in Table \ref{t1Q}. The values for $\langle Q\rangle$ are 
in general better localized at zero in case of the tempered runs. The results
for $\langle Q^2\rangle$ increase with lattice size, as is to be expected, and 
also for smaller $\kappa$. The errors of $\langle Q^2\rangle$ are rather large 
for the standard run and for the tempered run with 6 ensembles while for the 
tempered one with 7 ensembles they are moderate. This again indicates that
under the given conditions smaller spacing of $\kappa$ is more advantageous.

With respect to these errors one has to realize that within a tempering run 
the values found for $\langle Q^2\rangle$ at different $\kappa$ are correlated.
This phenomenon is well known from autocorrelation curves and mass 
determinations. To account for it in the fit to a corresponding
function the full weight matrix is to be used. To obtain this matrix 
fortunately one can rely on the fact that higher than two-point cumulants 
vanish and thus express four-point correlations by their disconnected parts 
\cite{pr89}. 

Since we do not intend to make a fit here it suffices to keep in mind that,
while the errors account properly for the individual values, they do not tell
about relative errors of neighboring values. In any case, better data from 
tempering, though correlated in $\kappa$, are to be preferred to poor ones 
from standard runs.

Concerning the magnitudes of $\langle Q^2\rangle$ in Table \ref{t1Q}, 
though errors are large, for the $12^4$ lattice some possible tendencies 
may be discussed. For large $\kappa$ the results of the tempered runs with 
6 and with 7 ensembles agree and have larger values than the standard run. 
Since the latter, as is obvious from Figure 2, only occasionally escapes from 
the value $Q=0$ this can be understood. On the other hand, for smaller $\kappa$
the values obtained in the run with 7 ensembles are below those of the other 
runs. In that case in Figure 2 the curve of standard HMC is seen to travel also
to larger $|Q|$, however, also to take quite some time to get back. A 
broadening of the distribution caused by this behavior explains the deviation.
Because of less accuracy and decorrelation, the run with 6 ensembles in its
behavior appears to be closer to the standard HMC case.

\section{Swap acceptance rates} \label{secBig}

With regard to large scale simulations of QCD performance predictions
are needed.  One potential problem of the tempering method has been
addressed in \cite{UKQCD}, namely the decrease of the swap acceptance
rate $\langle A\rangle$ with the lattice volume.  There the relation 
\begin{equation} \label{eqA}
\langle A \rangle = \mbox{erfc} \left( \frac{1}{2} 
\sqrt{\langle \Delta H \rangle} \right)
\end{equation}
has been used which has been shown in \cite{Bielefeld} to hold for
Metropolis-like algorithms. In \cite{UKQCD}, in simulations with 2 ensembles, 
the relation (\ref{eqA}) has been found to hold for a large range of 
$\langle \Delta H \rangle$. 

Our swap acceptance rates $\langle A \rangle$ for the $10^4$ and $12^4$ 
lattices here and for the $8^4$ lattice in \cite{us} are given in Table 
\ref{t1a}. To check the relation (\ref{eqA}) we have determined $y$ in
\begin{equation} \label{eqy}
\langle A \rangle = \mbox{erfc}(y)
\end{equation}
and have also listed the resulting quantities $100y/L^2$ and $10y/L$ for
our $L^4$ lattices in Table \ref{t1a} (the factors of 100 and 10 where
introduced to obtain numbers of $O(1)$). For linear scaling of
$\langle \Delta H \rangle$ with lattice volume by (\ref{eqA}) one
expects the quantity $100y/L^2$ to be constant. From Table \ref{t1a}
it is seen that this only holds for the case of 6 ensembles; for
the case with 7 ensembles, instead, the quantity $10y/L$ gets constant. 
In this context it should be remembered that the data from 7 ensembles,
i.e. from smaller $\Delta\kappa$, are our more accurate ones.

With respect to the data on the $8^4$ some caution is necessary because in
that case, at $\beta = 5.6$ and $0.15 \leq \kappa \leq 0.16$, the finite 
temperature phase transition \cite{PT} is crossed. This could influence
the scaling behavior of $\langle \Delta H \rangle$. However, in Table 
\ref{t1a} the $8^4$ data (within errors) fit into the trend of the other data. 
Thus, the small-volume phase transition does not affect our observation.

Clearly more ensembles will be needed on larger lattices if one wants to keep 
$\langle A\rangle$ and the parameter range constant. However, firstly, as our
example with 7 ensembles shows, for smaller $\Delta\kappa$ one does better than 
is expected from naive scaling arguments. Secondly, an open question is how the 
decrease of $\langle A \rangle$ and the slowing down of tunneling between
topological sectors compete. Within this respect one can hope that the 
speed-up more than compensates the additional effort by taking 
more ensembles. Thirdly, since results at several parameter values are 
needed anyway, having more points with less statistics is essentially 
equivalent to fewer points with more statistics.

\section{Discussion}

We have observed that parallel tempering considerably enhances the tunneling 
between different sectors of the topological charge. This enhancement also 
indicates an improvement of decorrelation for other observables. The method is 
particularly economical when several parameter values have to be studied 
in order to analyse the physical dependence. These features make parallel 
tempering an attractive method for QCD simulations. 

So far we have used equidistant parameter values. As known from simulated
tempering \cite{ker94,ker95} and from parallel tempering for spin glass 
\cite{HN} refinements using optimized distances are possible. In the present 
case such optimizations have been not practicable because of the amount of 
computer time which this would have needed. Similarly optimizing the parameter 
range, i.e.~the penetration into the range with easy tunneling, has not been 
feasible. The economics of computer time enforces to defer such optimizations
to the applications, i.e., ``to learn on the job''.

In large-scale QCD simulations the performance on larger lattices is important.
It has been seen here that definite predictions for this from smaller lattices 
are difficult. Nevertheless, for example, to make progress with the $\eta'$ 
problem, one should just try parallel tempering. Where data at several 
parameter values are needed, nothing can be lost doing the respective 
simulations simultaneously and making the parameters dynamical.

On the other hand, there are, of course, other problems in QCD for which the
method is attractive. Thus one might think of finite temperature studies, 
of phase structures with modified actions, or generally of questions where 
slowing down so far prevents from satisfactory solutions.

\section*{Acknowledgements}

The simulations were done on the CRAY T3E at
Konrad-Zuse-Zentrum f\"ur Informationstechnik Berlin.


\clearpage
\vspace*{20mm}

\begin{table}[h]
\caption{\label{t1}Values of $M_{Q}$ and $1/T_{Q}$ at $\beta = 5.6$.}
\vspace*{4mm}

\begin{tabular*}{16cm}{@{\extracolsep{\fill}}lllllll}

\hline
&
\multicolumn{2}{c}{standard HMC} &
\multicolumn{4}{c}{tempered HMC} \\

& & &
\multicolumn{2}{c}{6 ensembles} &
\multicolumn{2}{c}{7 ensembles} \\

& & &
\multicolumn{2}{c}{$\Delta\kappa = 0.0005$} &
\multicolumn{2}{c}{$\Delta\kappa = 0.00025$} \\

\cline{2-3}\cline{4-5}\cline{6-7}
&
\multicolumn{1}{c}{$10^4$} &
\multicolumn{1}{c}{$12^4$} &
\multicolumn{1}{c}{$10^4$} &
\multicolumn{1}{c}{$12^4$} &
\multicolumn{1}{c}{$10^4$} &
\multicolumn{1}{c}{$12^4$} \\

\hline\hline
\multicolumn{1}{c}{$\kappa$} & \multicolumn{6}{c}{$M_Q$} \\
\hline
0.15500 & 0.27(4)  & 0.47(4)  & 0.50(5)  & 0.77(5)  &          &         \\
0.15550 &          &          & 0.64(5)  & 0.90(8)  &          &         \\
0.15600 & 0.09(2)  & 0.30(4)  & 0.48(5)  & 0.80(8)  & 0.51(4)  & 0.91(5) \\
0.15625 &          &          &          &          & 0.57(4)  & 1.20(7) \\
0.15650 & 0.07(2)  & 0.33(3)  & 0.36(5)  & 0.75(7)  & 0.48(5)  & 1.17(7) \\
0.15675 &          &          &          &          & 0.38(4)  & 1.12(10)\\
0.15700 & 0.08(2)  & 0.20(5)  & 0.31(5)  & 0.53(7)  & 0.33(5)  & 0.96(9) \\
0.15725 &          &          &          &          & 0.23(5)  & 0.82(7) \\
0.15750 & 0.05(1)  & 0.09(2)  & 0.15(3)  & 0.28(4)  & 0.14(4)  & 0.51(7) \\

\hline\hline
\multicolumn{1}{c}{$\kappa$} & \multicolumn{6}{c}{$1/T_Q$} \\
\hline
0.15500 & 0.24(3)  & 0.38(2)  & 0.35(2)  & 0.48(2)  &          &         \\
0.15550 &          &          & 0.46(3)  & 0.53(2)  &          &         \\
0.15600 & 0.09(2)  & 0.27(3)  & 0.40(4)  & 0.45(3)  & 0.42(3)  & 0.55(2) \\
0.15625 &          &          &          &          & 0.48(3)  & 0.69(3) \\
0.15650 & 0.07(2)  & 0.28(2)  & 0.32(4)  & 0.47(2)  & 0.41(3)  & 0.69(2) \\
0.15675 &          &          &          &          & 0.32(3)  & 0.64(3) \\
0.15700 & 0.08(2)  & 0.18(4)  & 0.28(4)  & 0.36(3)  & 0.27(4)  & 0.61(3) \\
0.15725 &          &          &          &          & 0.21(4)  & 0.58(3) \\
0.15750 & 0.05(1)  & 0.09(2)  & 0.14(3)  & 0.22(3)  & 0.13(4)  & 0.36(4) \\
\hline

\end{tabular*}

\end{table}

\clearpage
\vspace*{20mm}

\begin{table}[h]
\caption{\label{t1Q}Values of $\langle Q\rangle$ and $\langle Q^2\rangle$
at $\beta = 5.6$.}
\vspace*{4mm}

\newlength{\X}
\settowidth{\X}{$-$}
\newcommand{\M}{\makebox[\X]{$-$}}  
\renewcommand{\P}{\makebox[\X]{}}   

\begin{tabular*}{16cm}{@{\extracolsep{\fill}}lllllll}

\hline
&
\multicolumn{2}{c}{standard HMC} &
\multicolumn{4}{c}{tempered HMC} \\

& & &
\multicolumn{2}{c}{6 ensembles} &
\multicolumn{2}{c}{7 ensembles} \\

& & &
\multicolumn{2}{c}{$\Delta\kappa = 0.0005$} &
\multicolumn{2}{c}{$\Delta\kappa = 0.00025$} \\

\cline{2-3}\cline{4-5}\cline{6-7}
&
\multicolumn{1}{c}{$10^4$} &
\multicolumn{1}{c}{$12^4$} &
\multicolumn{1}{c}{$10^4$} &
\multicolumn{1}{c}{$12^4$} &
\multicolumn{1}{c}{$10^4$} &
\multicolumn{1}{c}{$12^4$} \\

\hline\hline
\multicolumn{1}{c}{$\kappa$} & \multicolumn{6}{c}{$\langle Q \rangle$} \\
\hline

\hline
0.15500 &\P0.27(22) &\P0.01(19) &\P0.13(13) &\M0.16(22) &          &          \\
0.15550 &           &           &\P0.01(10) &\M0.05(14) &          &          \\
0.15600 &\P0.05(13) &\M0.26(32) &\M0.07(6)  &\M0.06(16) &\P0.13(8) &\P0.08(13)\\
0.15625 &           &           &           &           &\P0.08(6) &\P0.01(14)\\
0.15650 &\M0.15(8)  &\P0.15(26) &\M0.01(5)  &\M0.10(14) &\P0.05(4) &\M0.02(13)\\
0.15675 &           &           &           &           &\P0.02(3) &\P0.06(10)\\
0.15700 &\M0.04(9)  &\M0.16(16) &\M0.05(6)  &\M0.01(9)  &\P0.04(2) &\M0.01(8) \\
0.15725 &           &           &           &           &\P0.04(2) &\M0.09(5) \\
0.15750 &\P0.11(7)  &\P0.05(5)  &\P0.00(4)  &\P0.10(11) &\P0.04(2) &\M0.01(5) \\
\hline
\hline
\multicolumn{1}{c}{$\kappa$} & \multicolumn{6}{c}{$\langle Q^2 \rangle$} \\
\hline
0.15500 &\P1.48(27) &\P2.19(43) &\P1.14(20) &\P2.60(51) &          &          \\
0.15550 &           &           &\P0.81(12) &\P2.08(34) &          &          \\
0.15600 &\P0.45(19) &\P1.97(48) &\P0.46(6)  &\P2.06(35) &\P0.59(7) &\P1.64(14)\\
0.15625 &           &           &           &           &\P0.45(5) &\P1.56(13)\\
0.15650 &\P0.28(8)  &\P1.93(32) &\P0.34(5)  &\P1.79(38) &\P0.35(5) &\P1.51(15)\\
0.15675 &           &           &           &           &\P0.27(4) &\P1.36(17)\\
0.15700 &\P0.34(13) &\P0.90(32) &\P0.24(5)  &\P1.10(23) &\P0.22(4) &\P1.08(15)\\
0.15725 &           &           &           &           &\P0.14(3) &\P0.78(10)\\
0.15750 &\P0.17(6)  &\P0.21(5)  &\P0.14(4)  &\P0.56(8)  &\P0.10(3) &\P0.61(10)\\
\hline

\end{tabular*}

\end{table}

\clearpage
\vspace*{20mm}

\begin{table}[h]
\begin{center}
\caption{Swap acceptance data at $\beta = 5.6$. \label{t1a}}
\vspace*{4mm}
\begin{tabular*}{13cm}{@{\extracolsep{\fill}}l|llllll}

\hline
&
\multicolumn{3}{c}{6 ensembles} &
\multicolumn{3}{c}{7 ensembles} \\

&
\multicolumn{3}{c}{$0.155 \leq \kappa \leq 0.1575$} &
\multicolumn{3}{c}{$0.156 \leq \kappa \leq 0.1575$} \\

&
\multicolumn{3}{c}{$\Delta\kappa = 0.0005$} &
\multicolumn{3}{c}{$\Delta\kappa = 0.00025$} \\

\cline{2-4}\cline{5-7}

&
\multicolumn{1}{c}{$8^4$} &
\multicolumn{1}{c}{$10^4$} &
\multicolumn{1}{c}{$12^4$} &
\multicolumn{1}{c}{$8^4$} &
\multicolumn{1}{c}{$10^4$} &
\multicolumn{1}{c}{$12^4$} \\

\hline
\hline
$\langle A\rangle$
& \multicolumn{1}{c}{0.63(1)}
& \multicolumn{1}{c}{0.41(1)}
& \multicolumn{1}{c}{0.22(1)}
& \multicolumn{1}{c}{0.82(1)}
& \multicolumn{1}{c}{0.67(1)}
& \multicolumn{1}{c}{0.54(1)} \\

\hline
$100 y / L^2 $
&\multicolumn{1}{c}{0.98}
& \multicolumn{1}{c}{1.06}
& \multicolumn{1}{c}{0.99}
& \multicolumn{1}{c}{0.86}
& \multicolumn{1}{c}{0.72}
& \multicolumn{1}{c}{0.61} \\

\hline
$10 y / L $
&\multicolumn{1}{c}{0.95}
& \multicolumn{1}{c}{1.06}
& \multicolumn{1}{c}{1.18}
& \multicolumn{1}{c}{0.69}
& \multicolumn{1}{c}{0.72}
& \multicolumn{1}{c}{0.73} \\

\hline

\end{tabular*}
\end{center}
\end{table}

\clearpage
\begin{figure}[h]

\mbox{%
\qTEXT{Standard HMC}\hspace{\Width}%
\qTEXT{Tempered HMC}\hspace{\Width}%
\qTEXT{Tempered HMC}}

\mbox{%
\qtext{}\hspace{\Width}%
\qtext{6 ensembles}\hspace{\Width}%
\qtext{7 ensembles}}

\mbox{%
\qtext{}\hspace{\Width}%
\qtext{$0.155 \leq \kappa \leq 0.1575$}\hspace{\Width}%
\qtext{$0.156 \leq \kappa \leq 0.1575$}}

\mbox{%
\qtext{}\hspace{\Width}%
\qtext{$\Delta\kappa = 0.0005$}\hspace{\Width}%
\qtext{$\Delta\kappa = 0.00025$}}

\bigskip

\mbox{\qpic{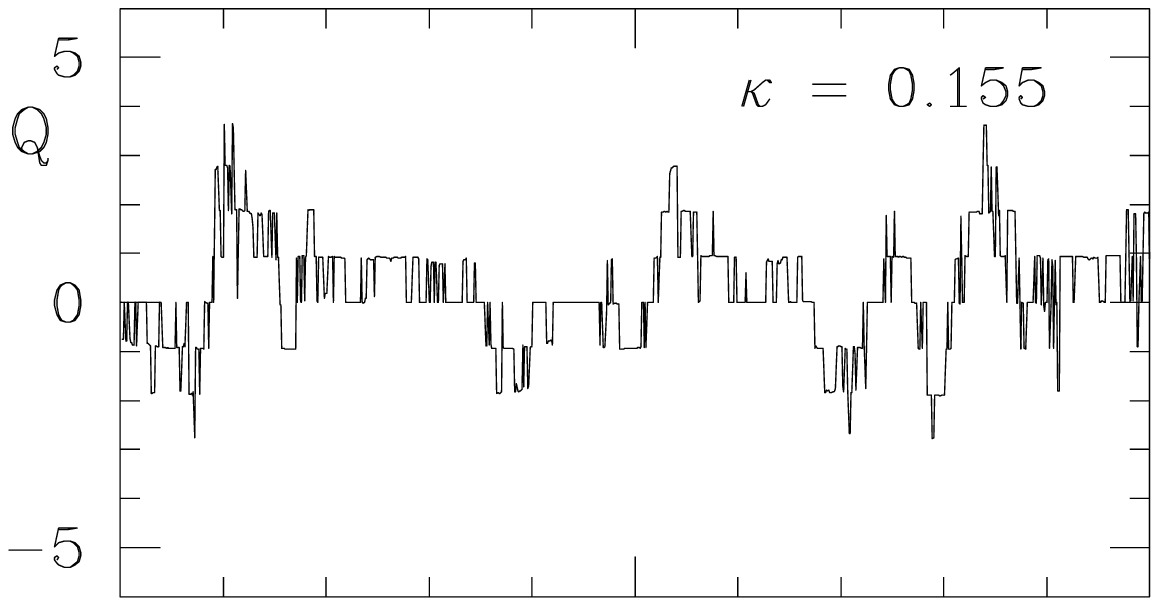}\hspace{\Width}\qpic{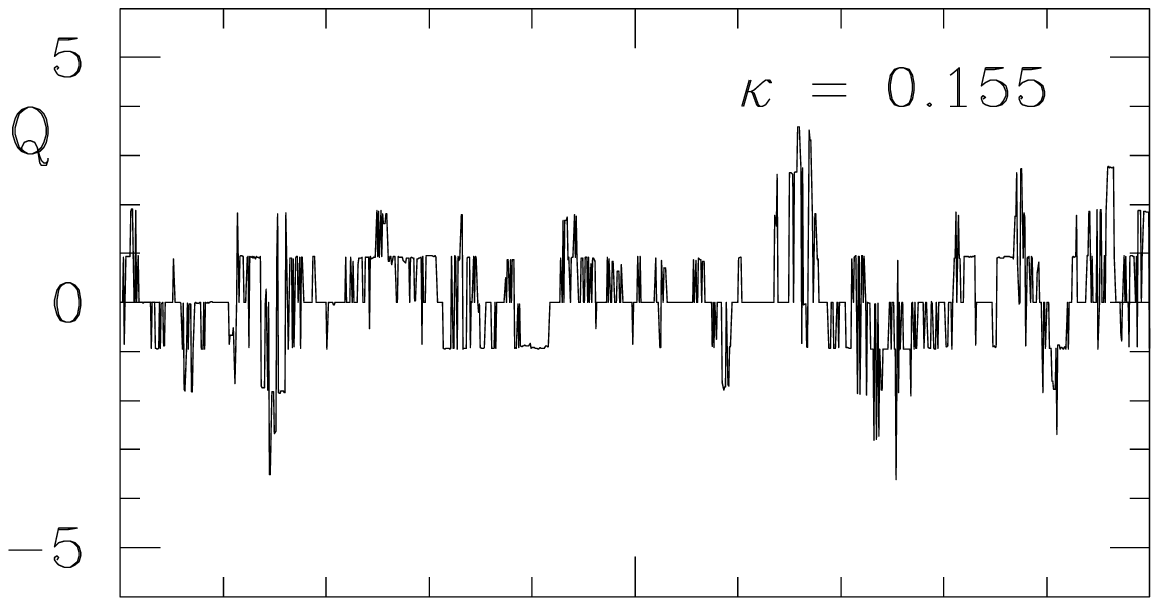}}
\mbox{\qpic{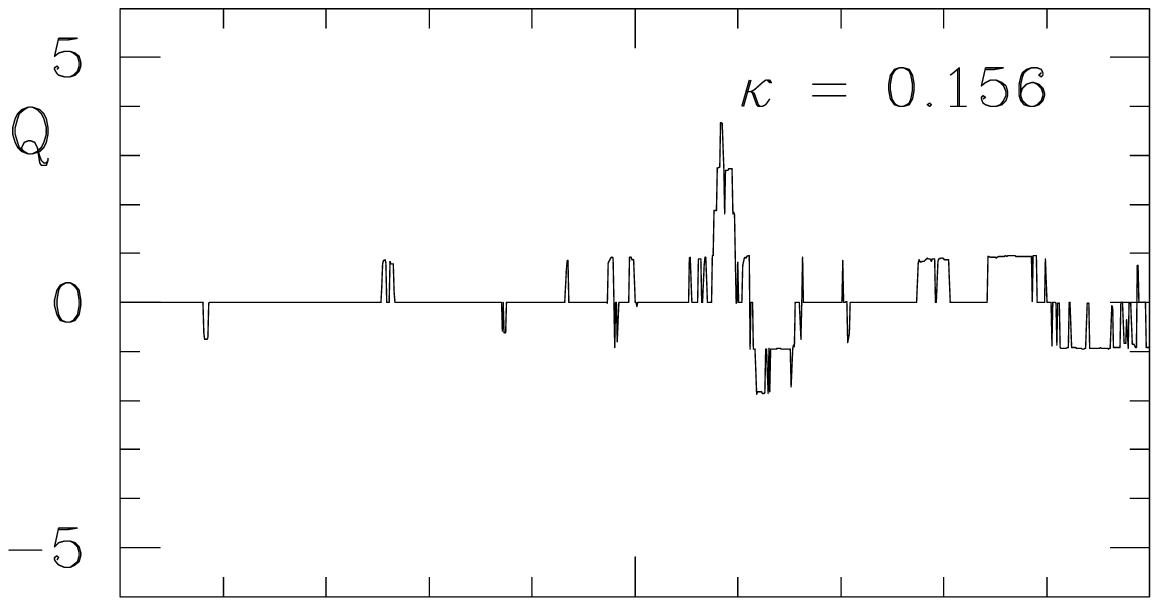}\hspace{\Width}\qpic{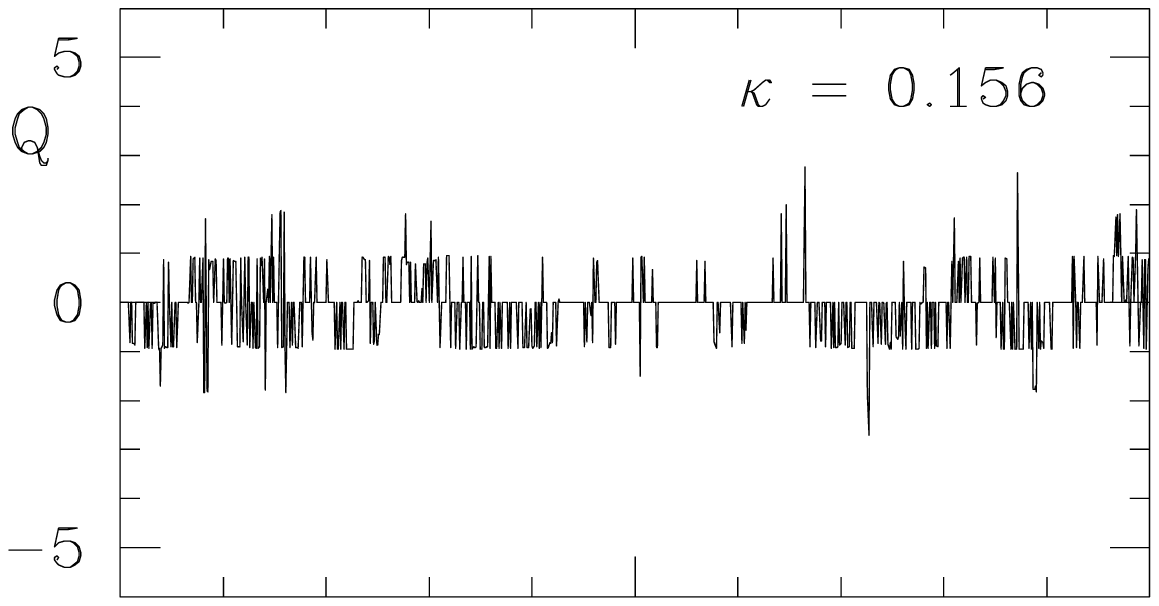}\hspace{\Width}\qpic{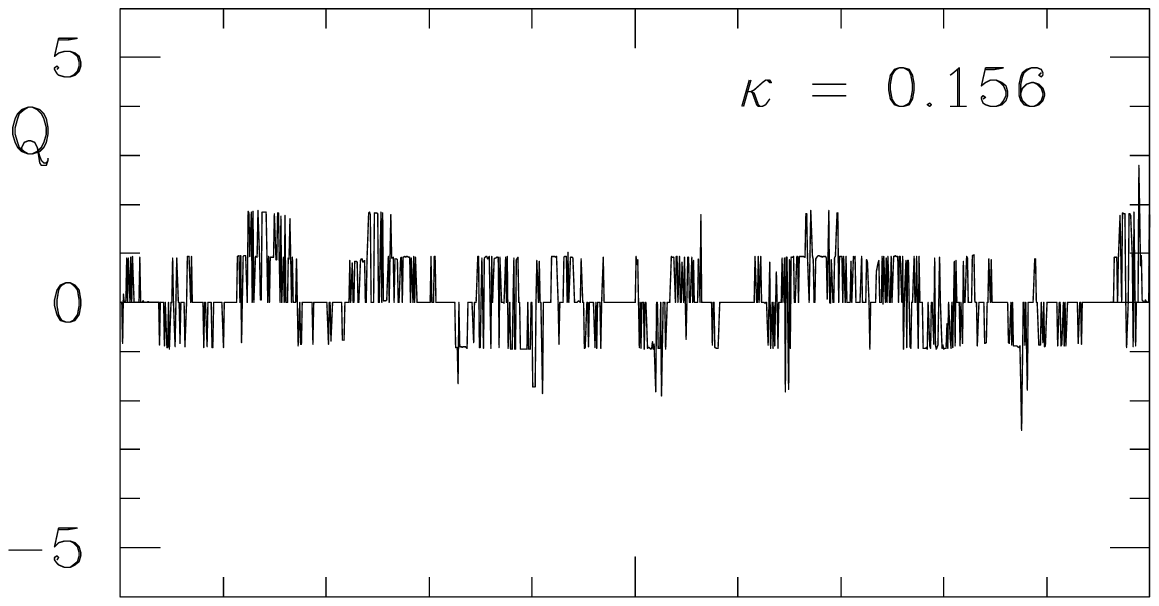}}
\mbox{\qpic{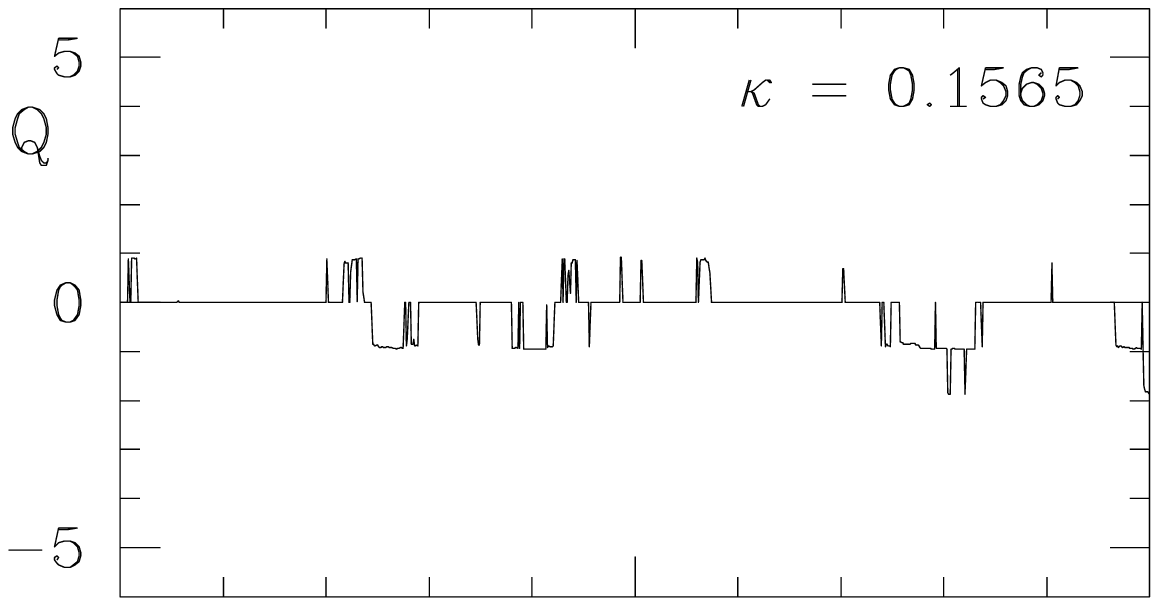}\hspace{\Width}\qpic{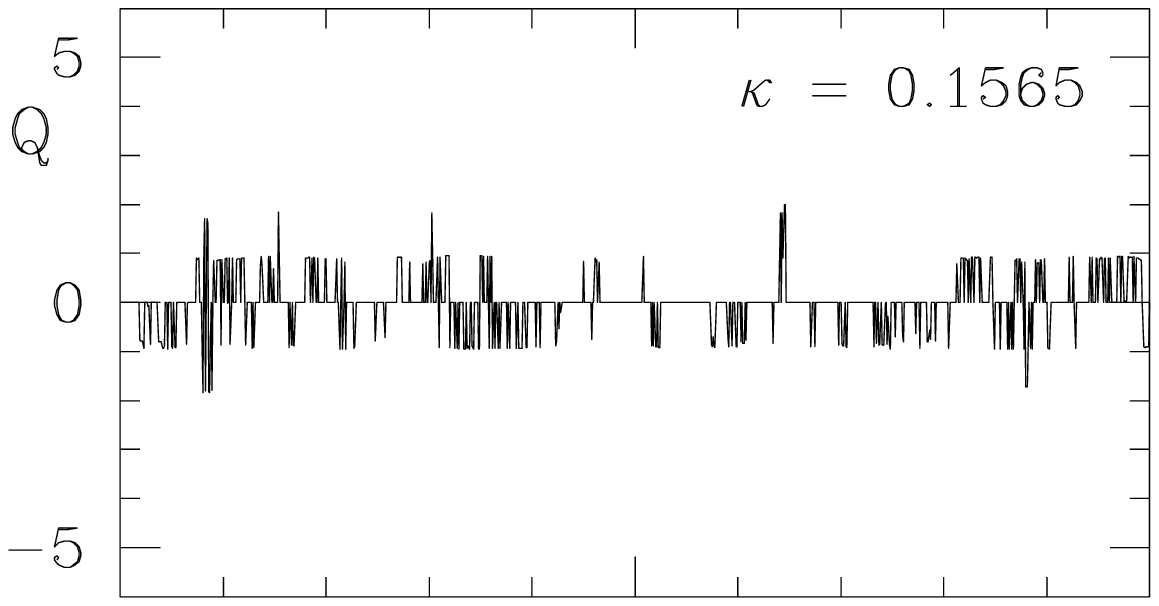}\hspace{\Width}\qpic{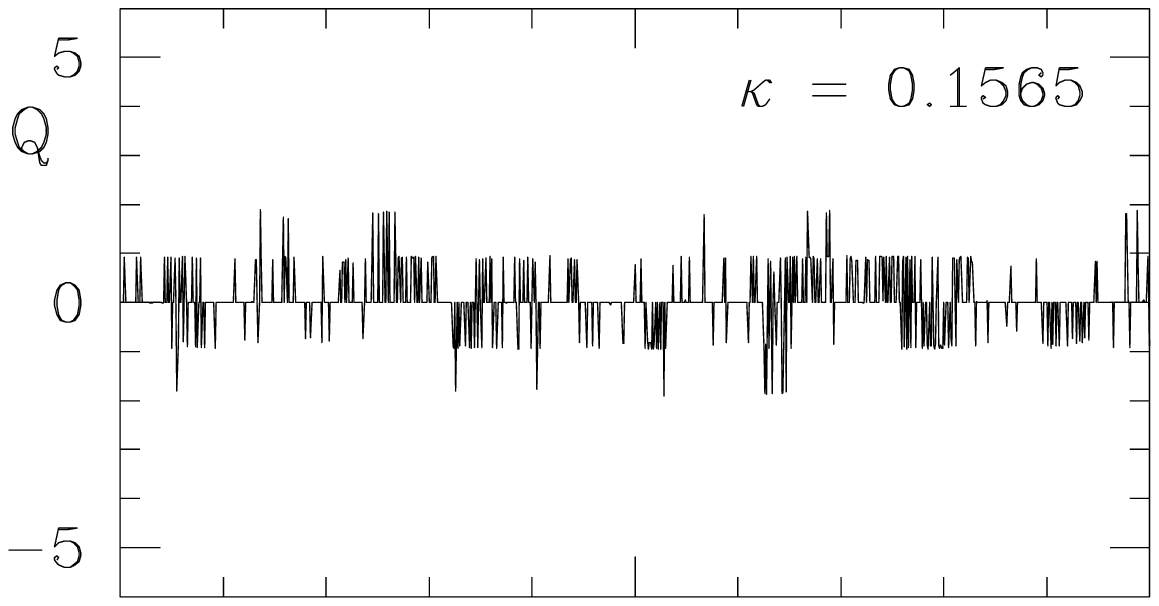}}
\mbox{\qpic{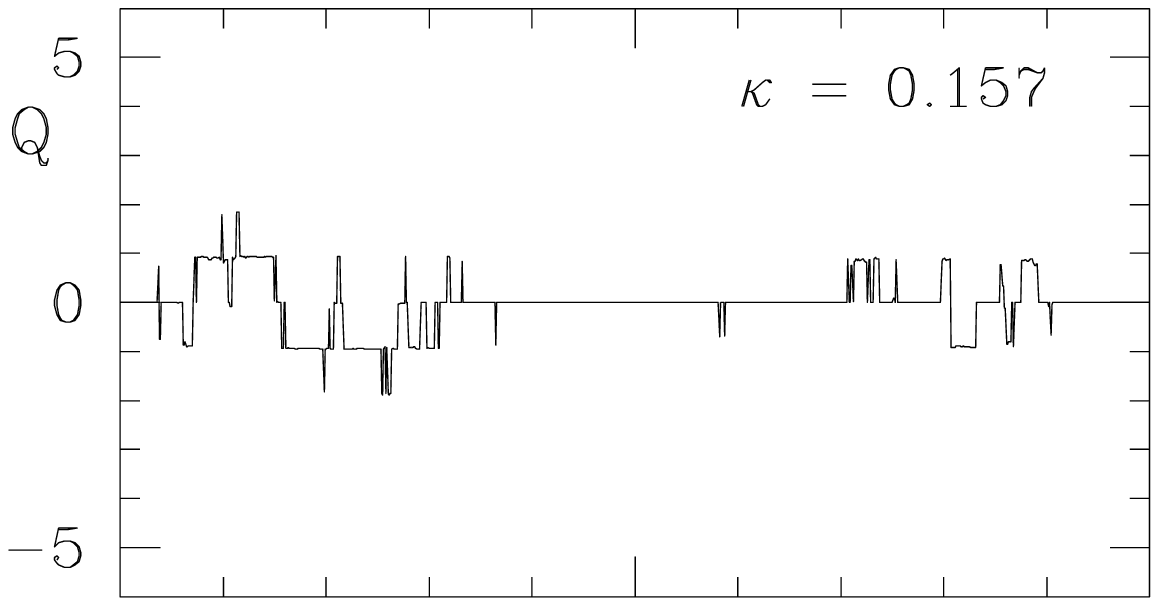}\hspace{\Width}\qpic{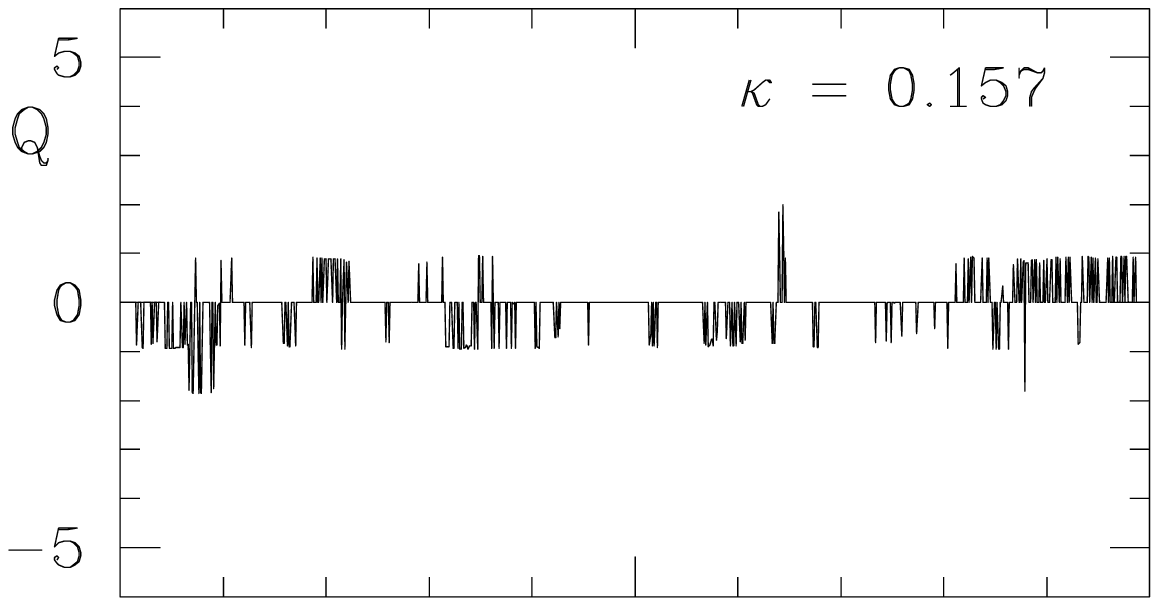}\hspace{\Width}\qpic{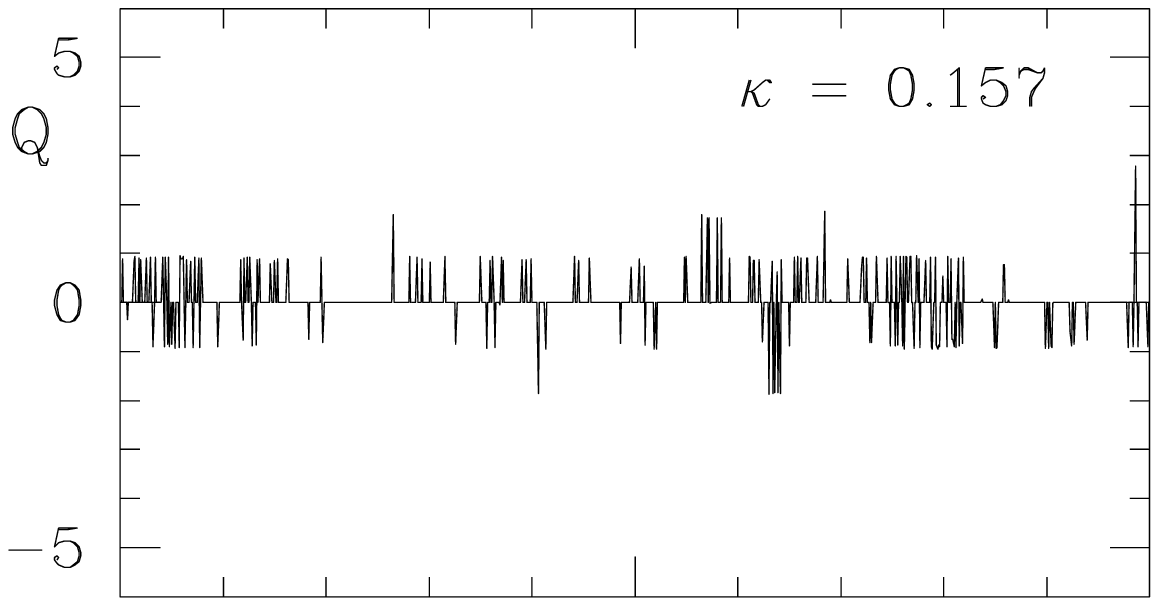}}
\mbox{\qpic{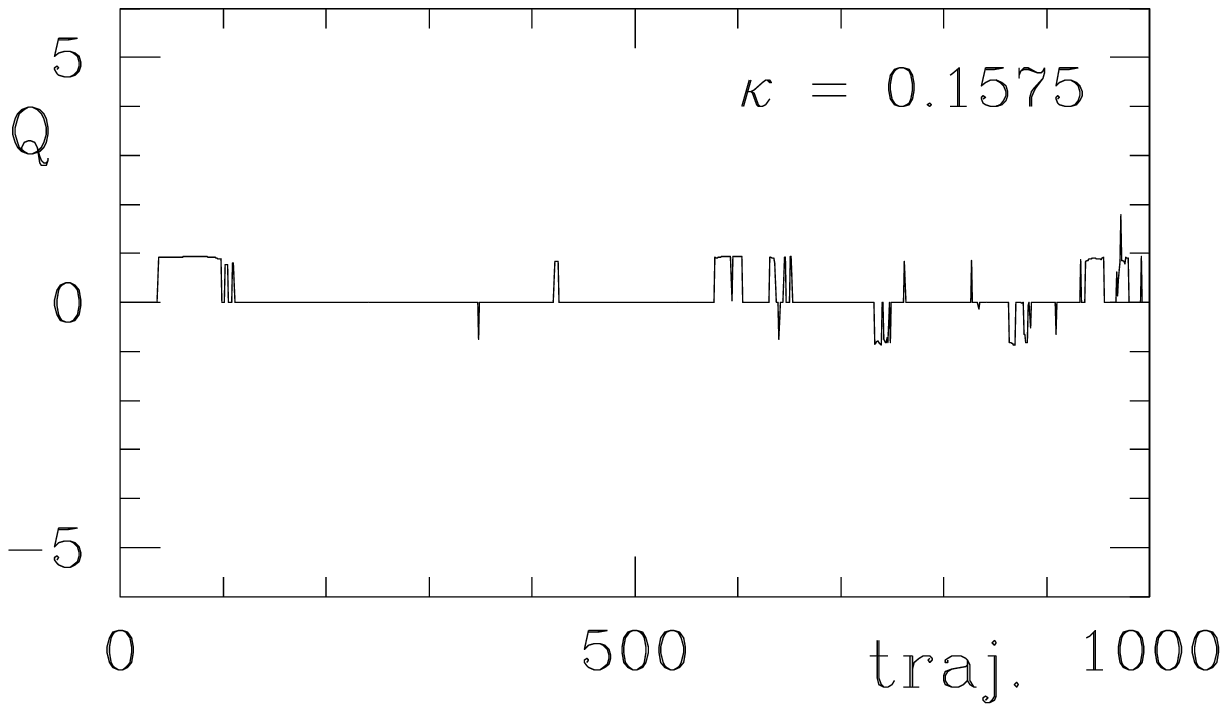}\hspace{\Width}\qpic{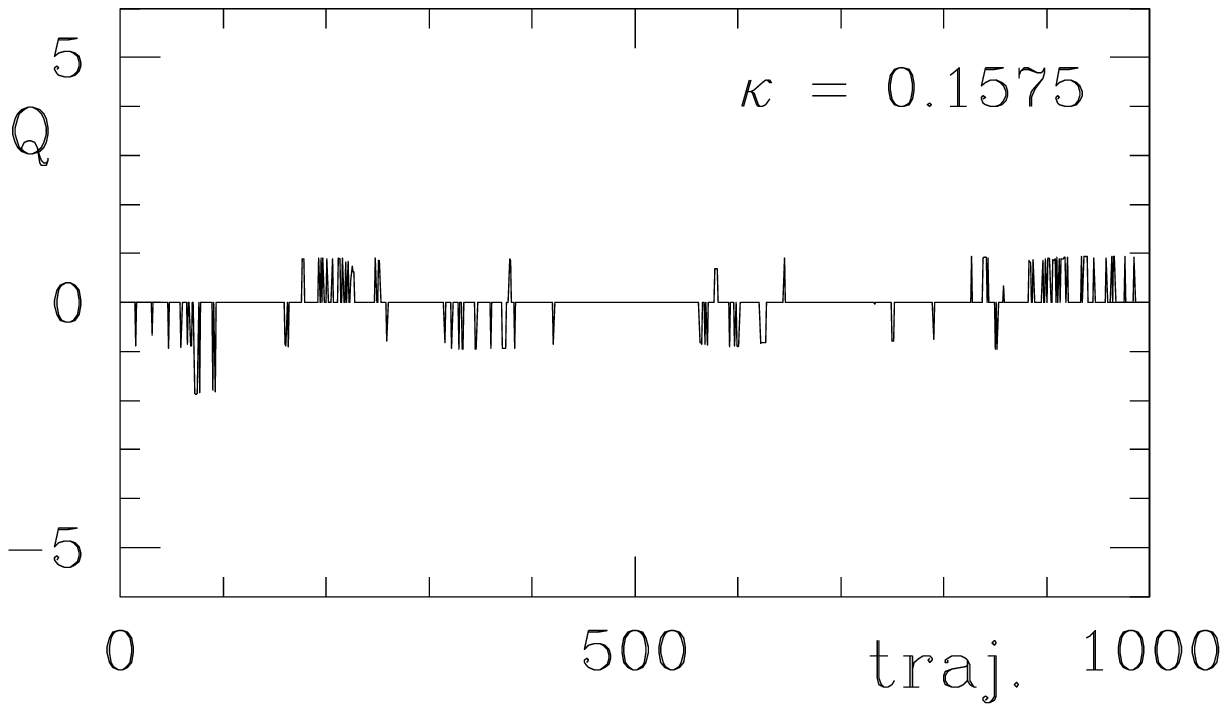}\hspace{\Width}\qpic{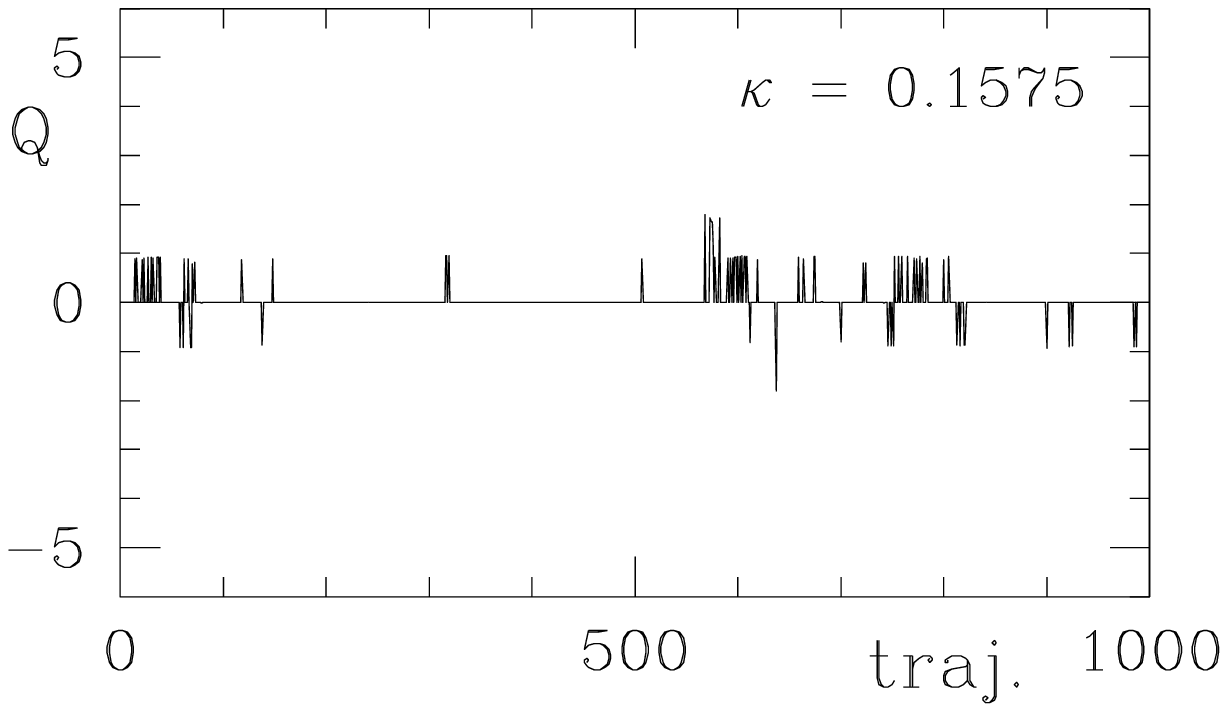}}

\bigskip

\caption{Time series of $Q$ for standard and tempered HMC 
on  $10^4$ lattice at $\beta = 5.6$.  The series for some intermediate 
$\kappa$-values are not shown (see Table \ref{t1} for full list).}
\end{figure}

\clearpage
\begin{figure}[h]

\mbox{%
\qTEXT{Standard HMC}\hspace{\Width}%
\qTEXT{Tempered HMC}\hspace{\Width}%
\qTEXT{Tempered HMC}}

\mbox{%
\qtext{}\hspace{\Width}%
\qtext{6 ensembles}\hspace{\Width}%
\qtext{7 ensembles}}

\mbox{%
\qtext{}\hspace{\Width}%
\qtext{$0.155 \leq \kappa \leq 0.1575$}\hspace{\Width}%
\qtext{$0.156 \leq \kappa \leq 0.1575$}}

\mbox{%
\qtext{}\hspace{\Width}%
\qtext{$\Delta\kappa = 0.0005$}\hspace{\Width}%
\qtext{$\Delta\kappa = 0.00025$}}

\bigskip

\mbox{\qpic{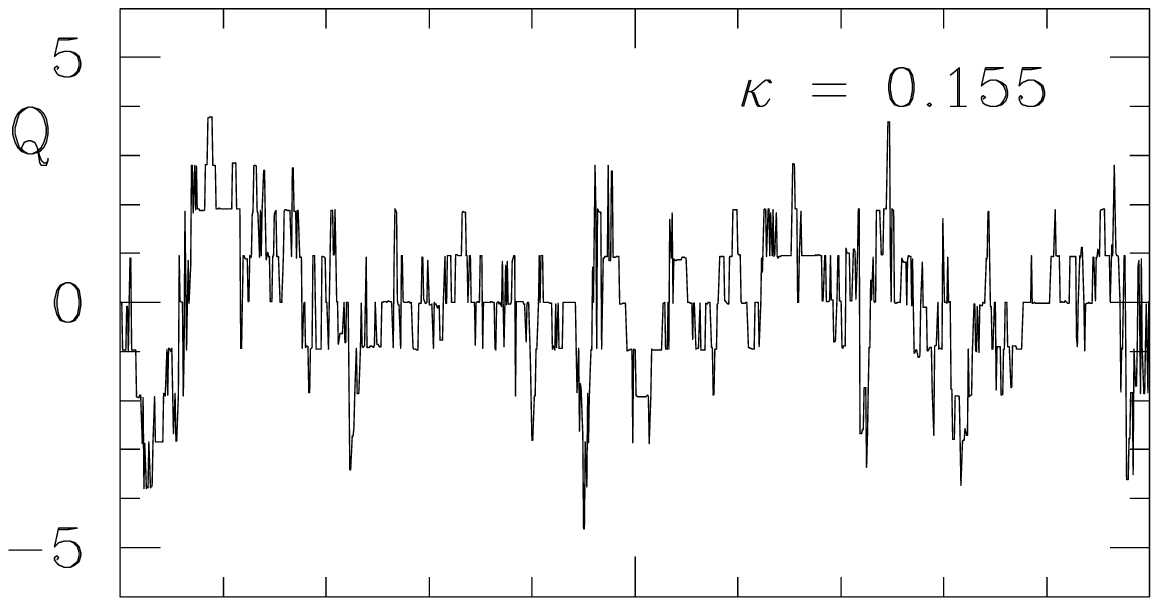}\hspace{\Width}\qpic{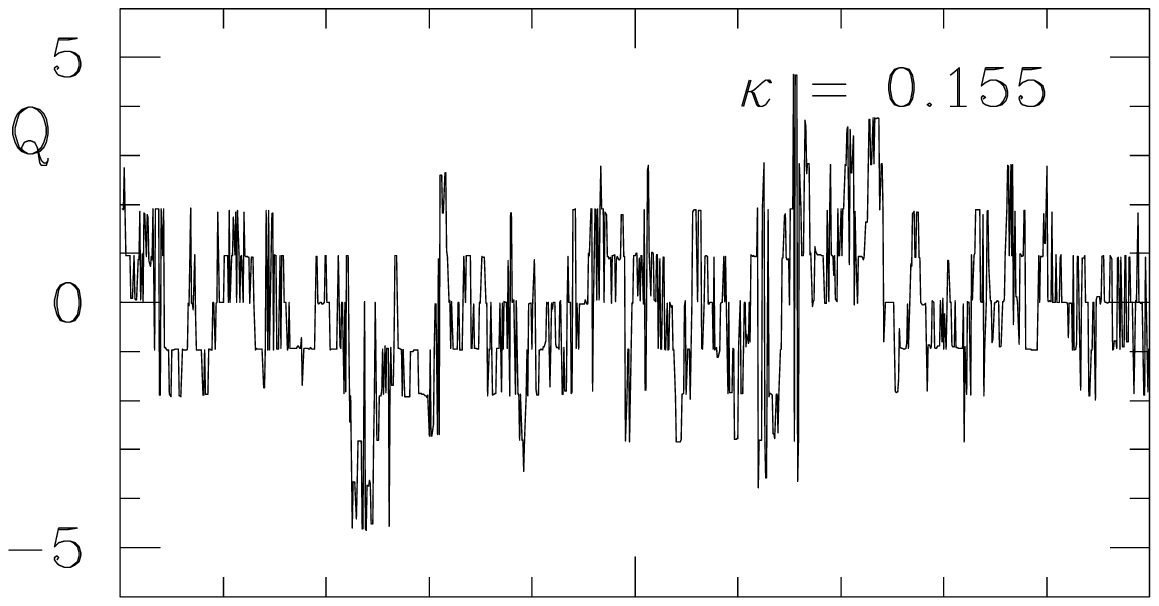}}
\mbox{\qpic{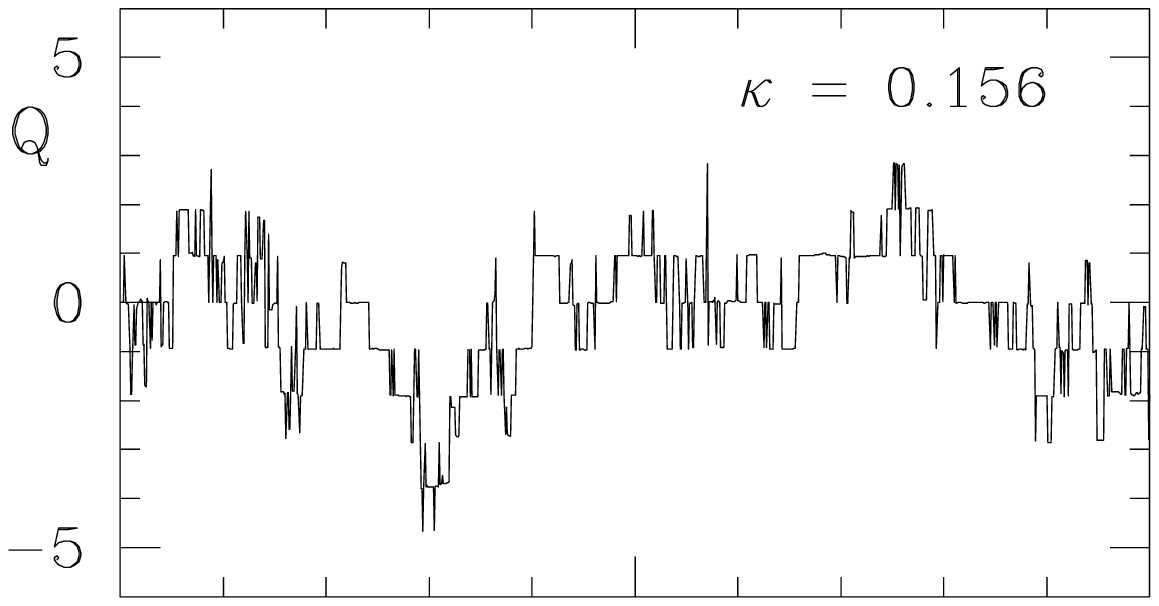}\hspace{\Width}\qpic{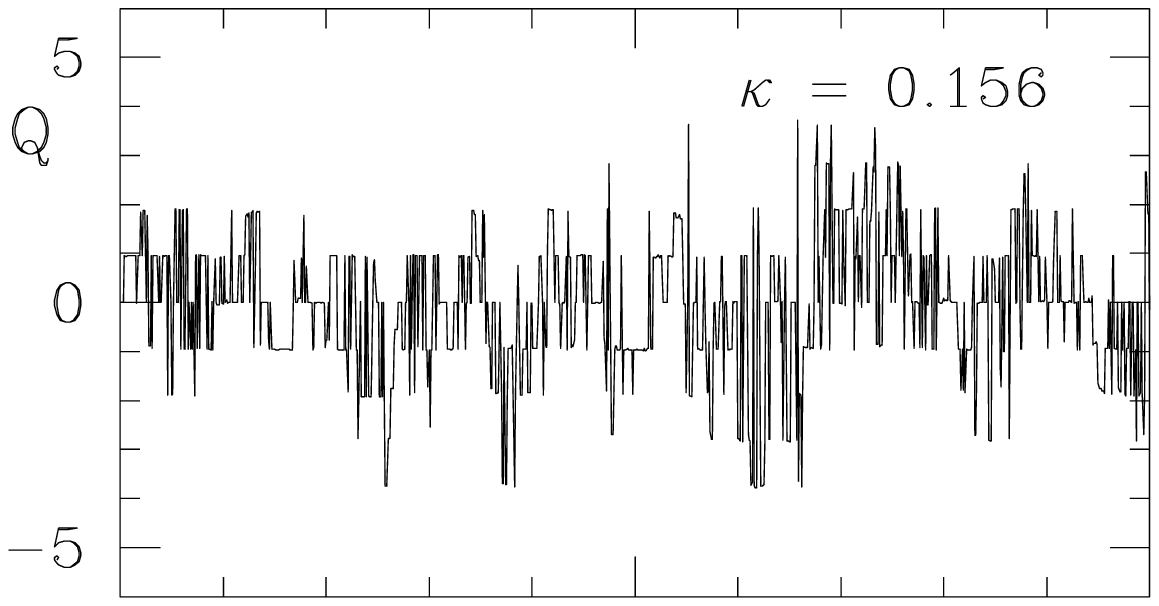}\hspace{\Width}\qpic{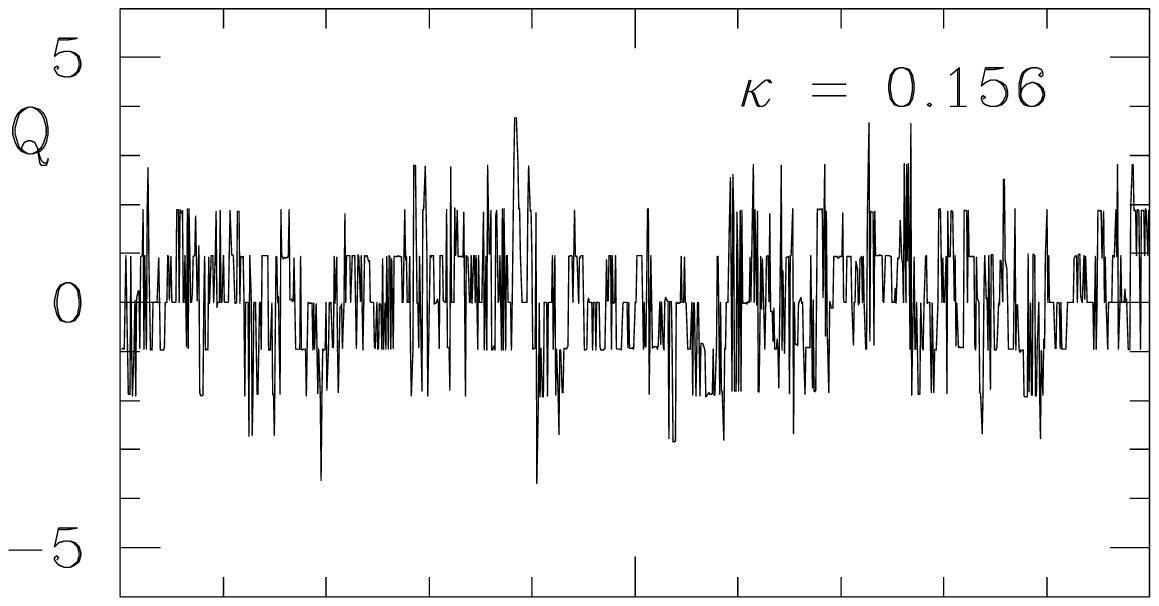}}
\mbox{\qpic{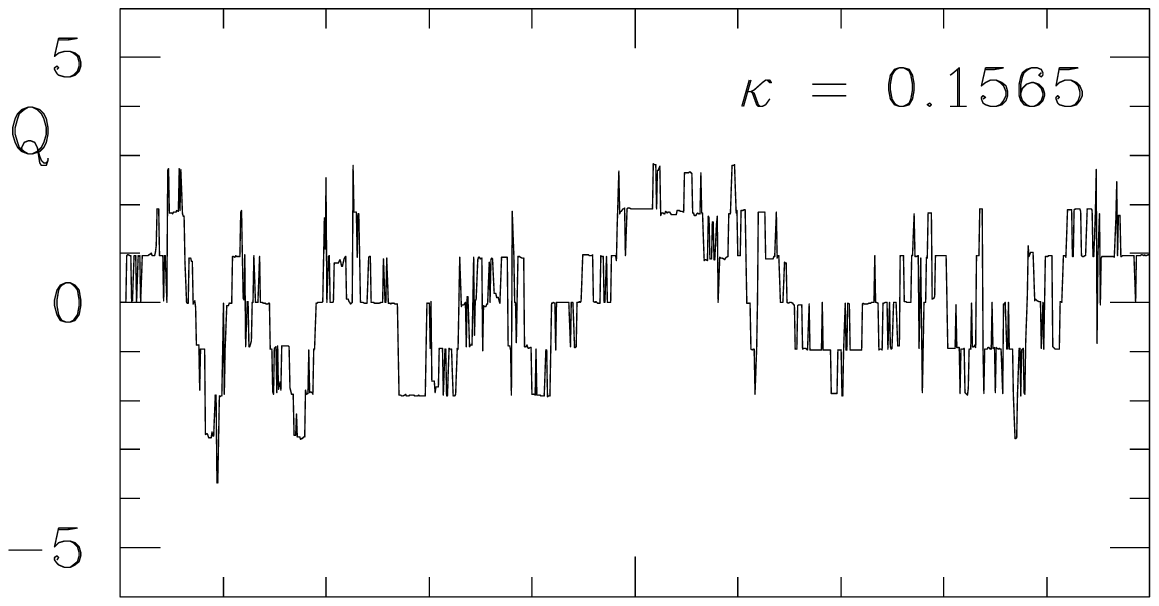}\hspace{\Width}\qpic{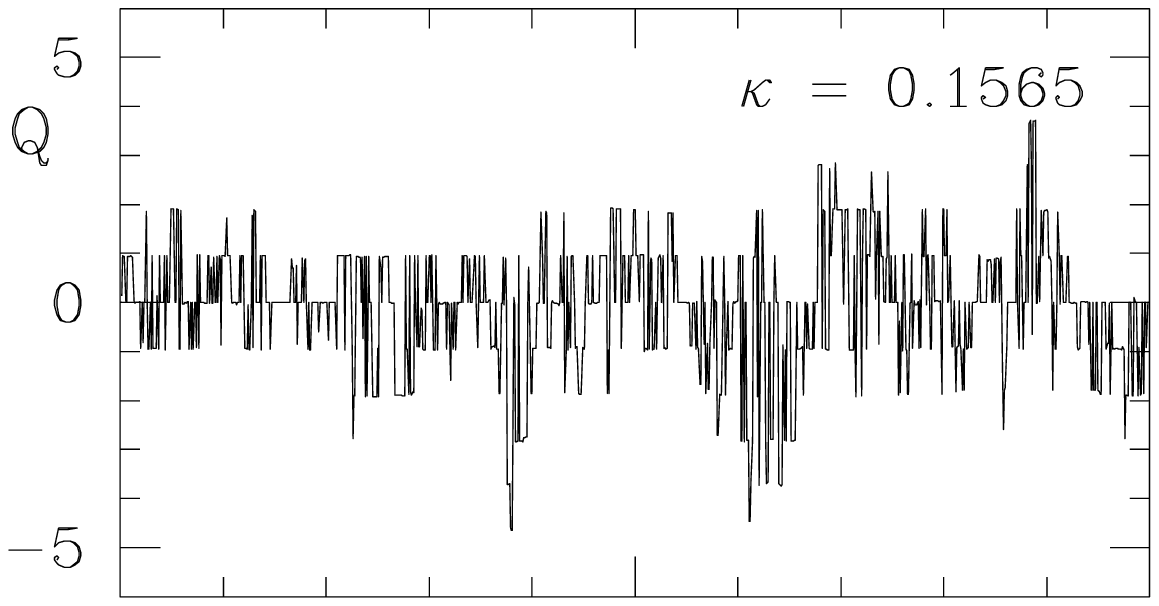}\hspace{\Width}\qpic{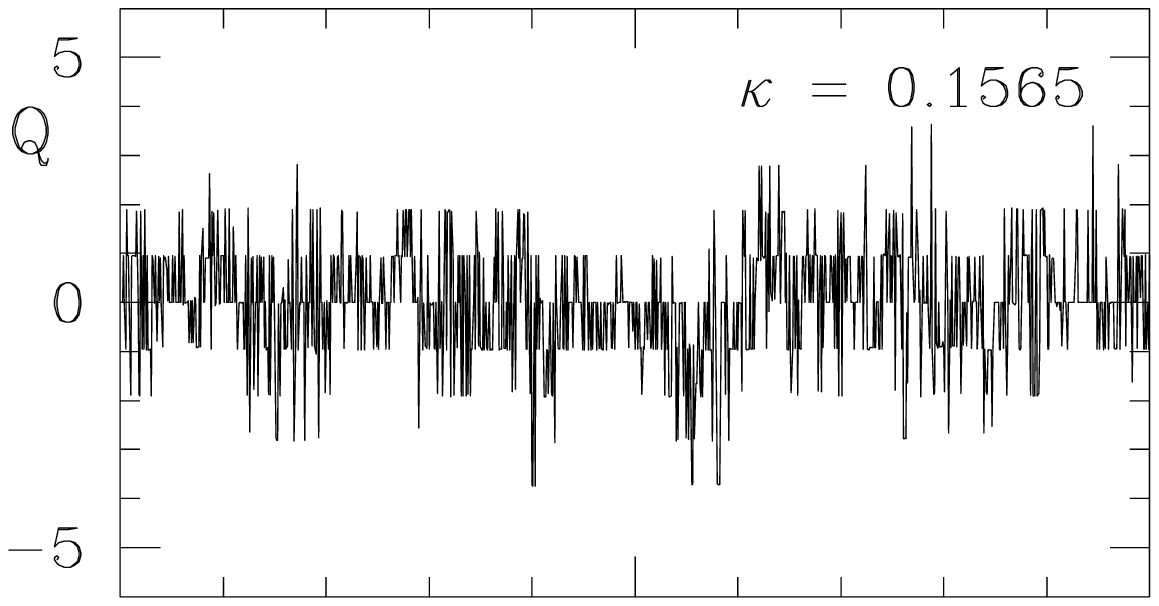}}
\mbox{\qpic{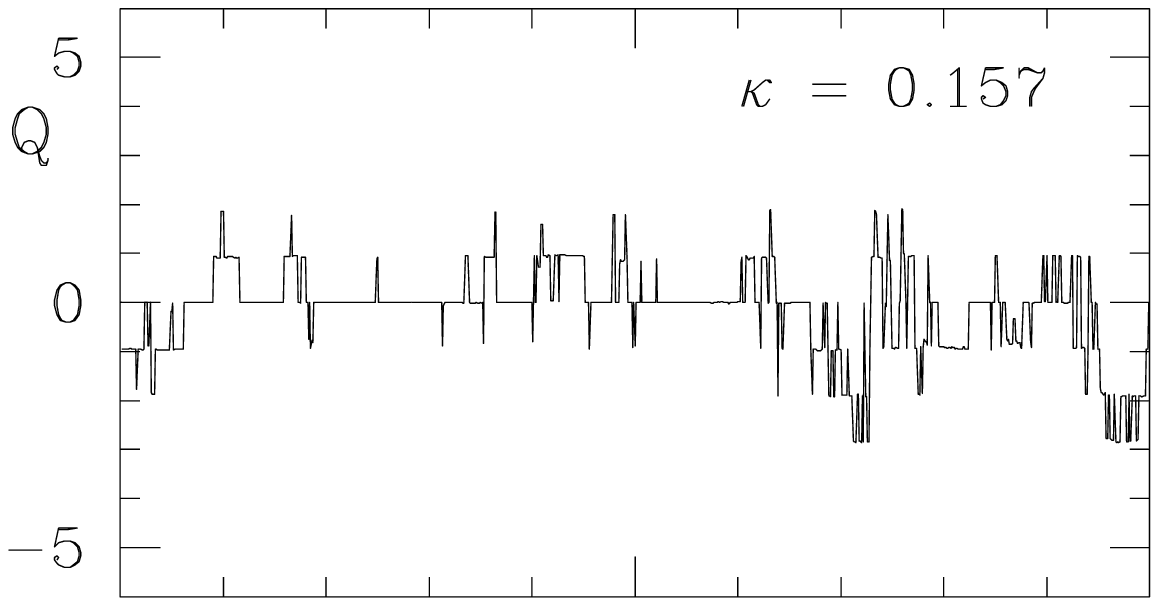}\hspace{\Width}\qpic{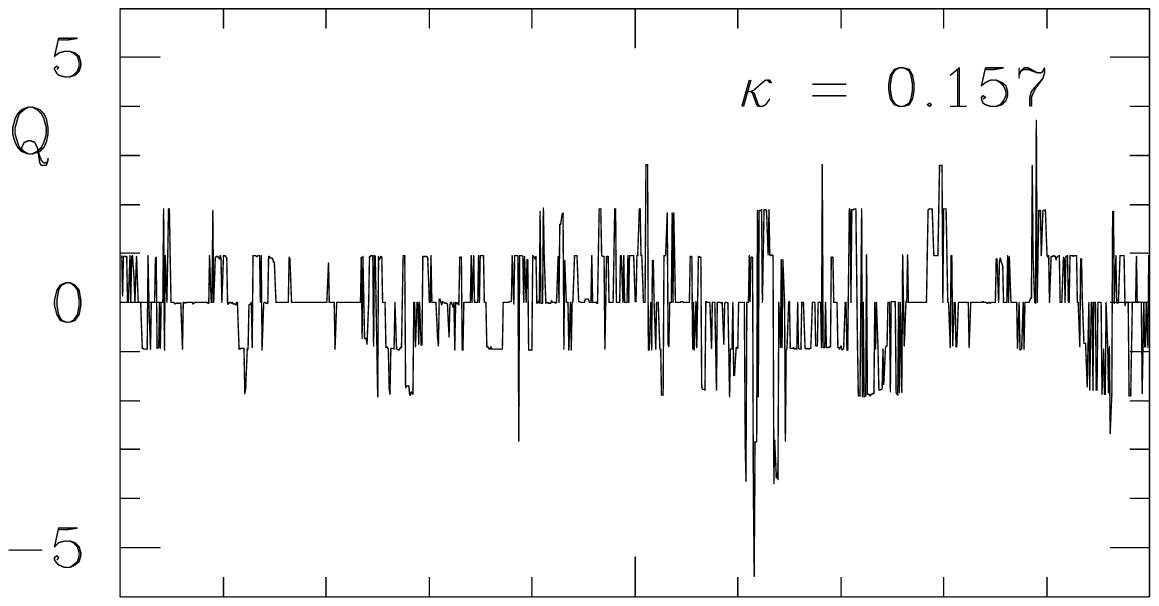}\hspace{\Width}\qpic{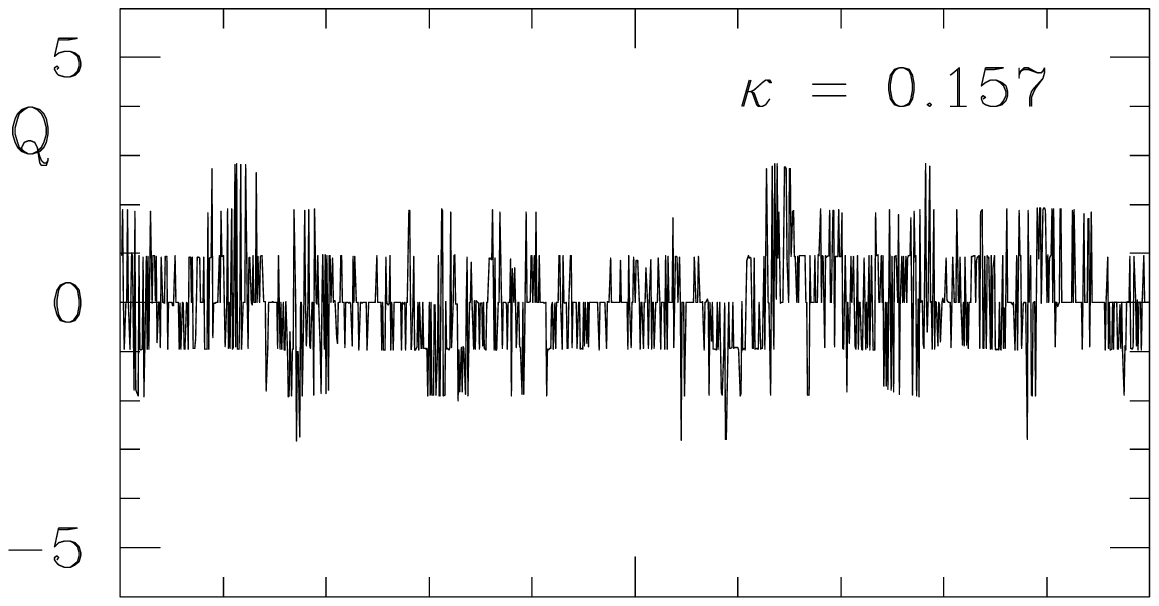}}
\mbox{\qpic{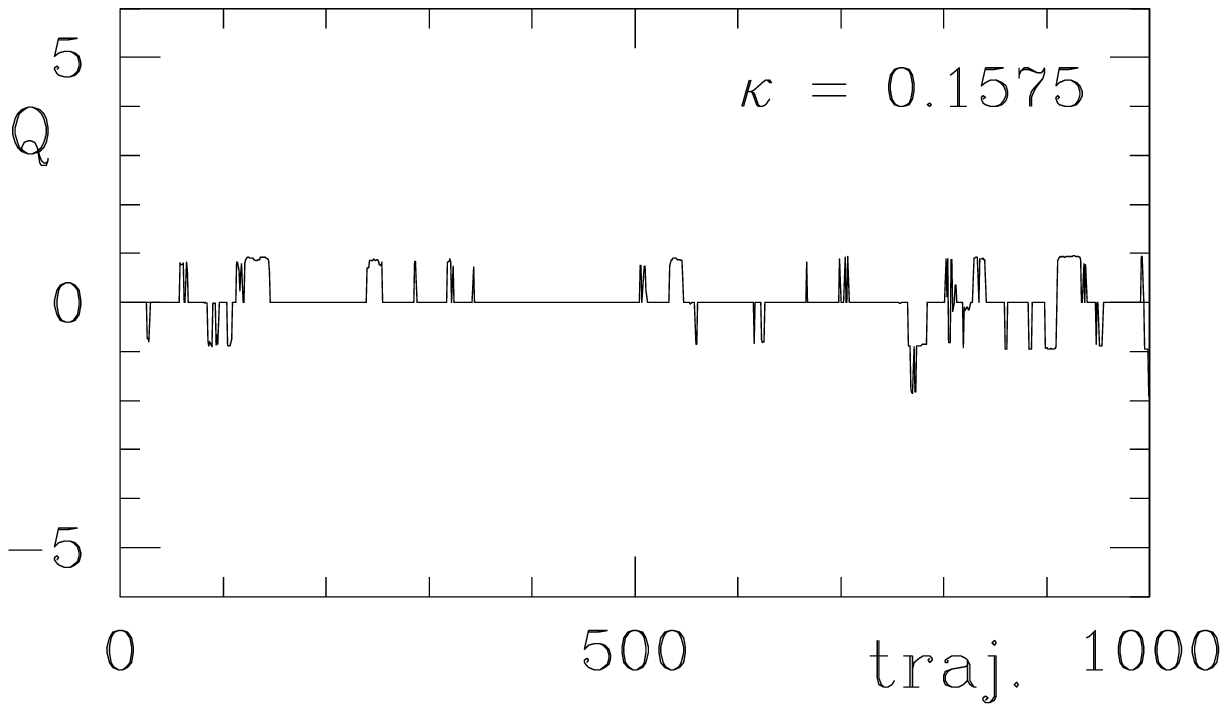}\hspace{\Width}\qpic{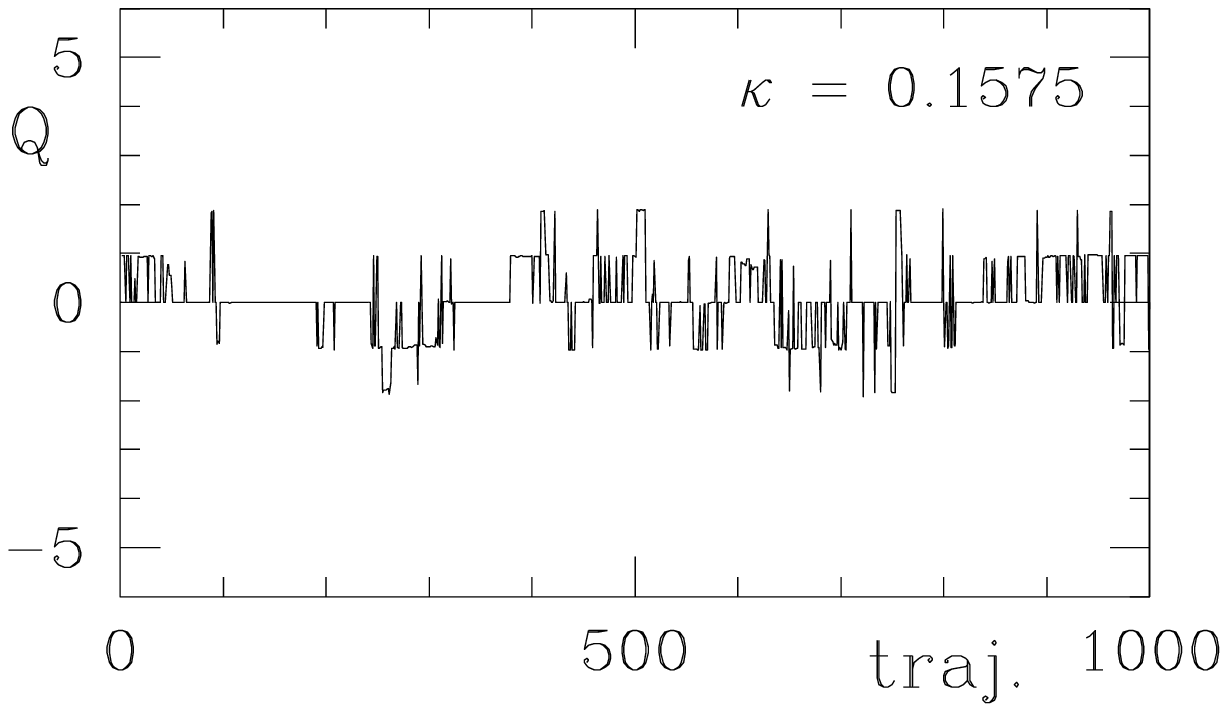}\hspace{\Width}\qpic{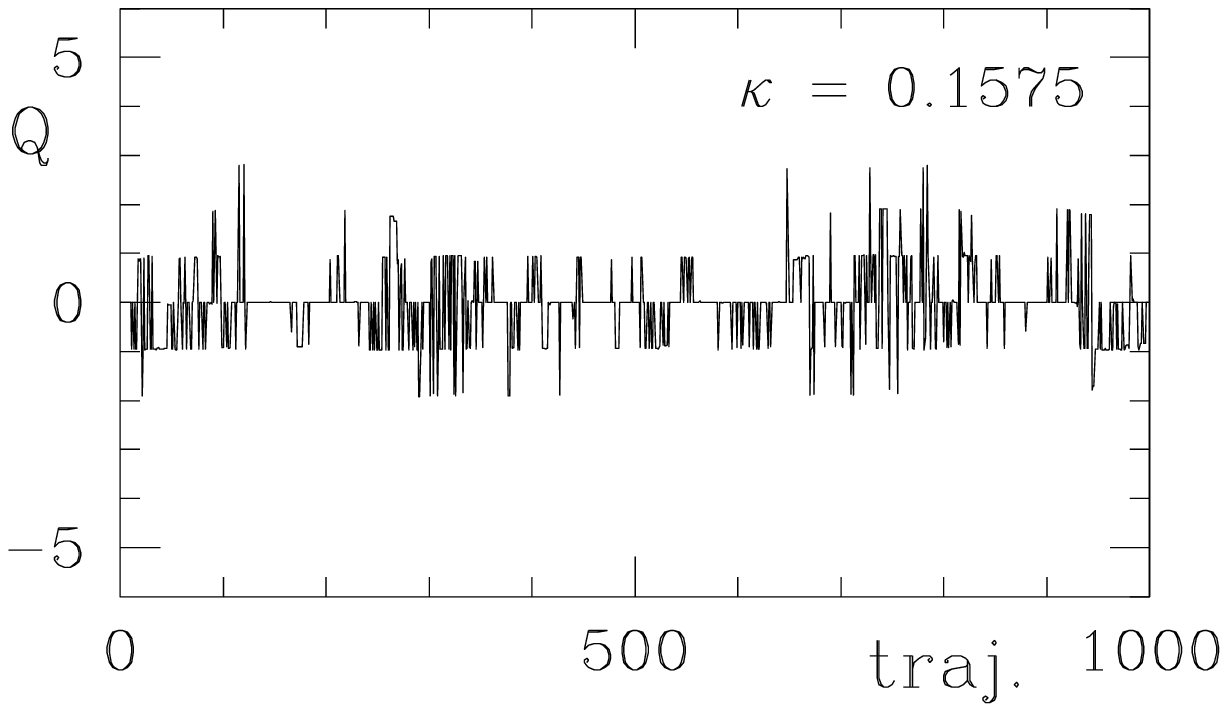}}

\bigskip

\caption{Time series of $Q$ for standard and tempered HMC 
on  $12^4$ lattice at $\beta = 5.6$.  The series for some intermediate 
$\kappa$-values are not shown (see Table \ref{t1} for full list).}
\end{figure}


\begin{thebibliography}{99}
\frenchspacing

\bibitem{MMP}  
M.~M\"uller-Preussker, Proc. of the XXVI Int. Conf. on
High Energy Physics, Dallas, Texas (1992), 1545.

\bibitem{Pisa}  
B.~All\'es, G.~Boyd, M.~D'Elia, A.~Di~Giacomo and E.~Vicari,
Phys. Lett. B 359 (1996) 107.

\bibitem{SESAM}
B.~All\'es, G.~Bali, M.~D'Elia, A.~Di~Giacomo, N.~Eicker,
K.~Schilling, A.~Spitz, S.~G\"usken, H.~Hoeber, Th.~Lippert,
T.~Struckmann, P.~Ueberholz and J.~Viehoff,
Phys. Rev. D 58 (1998) 071503.

\bibitem{CP-PACS}   
A.~Ali~Khan, S.~Aoki, R.~Burkhalter, S.~Ejiri, M.~Fukugita,
S.~Hashimoto, N.~Ishizuka, Y.~Iwasaki, K.~Kanaya, T.~Kaneko,
Y.~Kuramashi, T.~Manke, K.~Nagai, M.~Okawa, H.P.~Shanahan, A.~Ukawa
and T.~Yoshi\'e,
Nucl. Phys. B. (Proc. Suppl.) 83--84 (2000) 162.

\bibitem{SESAM2} 
T.~Struckmann, K.~Schilling, P.~Ueberholz, N.~Eicker, S.~G\"usken,
Th.~Lippert, H.~Neff, B.~Orth and J.~Viehoff,
\texttt{hep-lat/0006012}.

\bibitem{mar92}E. Marinari und G. Parisi, Europhys. Lett. 19 (1992) 451.
\bibitem{ker93}W. Kerler und A. Weber, Phys. Rev. B 47 (1993) R11563.
\bibitem{ker94}W. Kerler und P. Rehberg, Phys. Rev. E 50 (1994) 4220.
\bibitem{ker95}W. Kerler, C. Rebbi und A. Weber, Phys. Rev. D 50 (1994) 6984;
                  Nucl. Phys. B 450 (1995) 452.
\bibitem{boy97}G. Boyd, \texttt{hep-lat/9701009}.
 
\bibitem{HN}
K.~Hukushima and K.~Nemoto, \texttt{cond-mat/9512035}.

\bibitem{Boyd}
G.~Boyd, Nucl. Phys. B (Proc. Suppl.) 60A (1998) 341.

\bibitem{UKQCD} 
B.~Jo\'o,  B.~Pendleton, S.M.~Pickles, Z.~Sroczynski, A.C.~Irving,
and J.C.~Sexton,
Phys. Rev. D 59 (1999) 114501.

\bibitem{us}
E.-M.~Ilgenfritz, W.~Kerler and H.~St\"uben, 
Nucl. Phys. B. (Proc. Suppl.) 83--84 (2000) 831.
H.~St\"uben in \emph{Lattice Fermions and Structure of the Vacuum},
V.~Mitrjushkin and G.~Schierholz (eds.), (Kluwer Academic Publishers,
2000) p.~211.

\bibitem{EM}
For a review see: E.~Marinari, \texttt{cond-mat/9612010}.  

\bibitem{pr89} M.B. Priestley, {\it Spectral analysis and time series}
               (Academic Press, London, 1998).

\bibitem{Bielefeld}
S.~Gupta, A.~Irb\"ack, F.~Karsch and B.~Peterson,
Phys. Lett. B 242 (1990) 437.

\bibitem{PT}
Y.~Iwasaki, K.~Kanaya, S.~Kaya, S.~Sakai and T.~Yoshi\'e,
Phys. Rev. D 54 (1996) 7010.

\end{thebibliography}
\end{document}